\documentclass[11pt,a4paper]{article}
\pdfoutput=1
\usepackage{jcappub}
\newcommand{\figh}[2]{\includegraphics[height=#1\textheight]{#2}}

\newcommand{\uhecr}{\mbox{UHECR}}
\newcommand{\uhecrs}{\mbox{UHECRs}}

\newcommand{\Nsrc}{\ensuremath{N_\mathrm{src}}}
\newcommand{\Nside}{\ensuremath{N_\mathrm{side}}}

\newcommand{\Xmax}{\ensuremath{X_\mathrm{max}}}

\begin{document}

\title{%
	 A study of an energy-dependent anisotropy of cosmic rays beyond the GZK
	 cut-off with deep neural networks
	 }

\author[a,b]{Oleg Kalashev,}
\author[a,c,d]{Maxim~Pshirkov,}
\author[e]{Mikhail~Zotov}

\affiliation[a]{Institute for Nuclear Research of the Russian Academy of
	Sciences, Moscow, 117312, Russia}
\affiliation[b]{Moscow Institute for Physics and Technology, 9 Institutskiy per.,
	Dolgoprudny, Moscow Region, 141701 Russia}
\affiliation[c]{Sternberg Astronomical Institute, Lomonosov Moscow State
	University, Moscow, 119992, Russia}
\affiliation[d]{Lebedev Physical Institute, Pushchino Radio
	Astronomy Observatory, 142290, Russia}
\affiliation[e]{Skobeltsyn Institute of Nuclear Physics,
	Lomonosov Moscow State University, Moscow, 119991, Russia}

\abstract{%
    In this letter, we present an update of a method for analysing arrival directions of
	ultra-high-energy cosmic rays (UHECRs) above the Greisen--Zatsepin--Kuz'min cut-off with
	a deep convolutional neural network developed originally in Kalashev,
	Pshirkov, Zotov (2020).
	Namely, we introduce energy as another variable employed in the analysis.
	This allows us to take into account the intrinsic uncertainties in energy of primary cosmic
	rays present in any experiment, which were not taken into account in the previous study,
	without any loss of quality of the classifier.
	We present the architecture of the new neural network, results of its application
	to mock maps of UHECR arrival directions and outline possible directions of a
	further improvement of the method.
	}

\emailAdd{kalashev@inr.ac.ru}
\emailAdd{pshirkov@sai.msu.ru}
\emailAdd{zotov@eas.sinp.msu.ru}

\keywords{ultra-high-energy cosmic rays, anisotropy,
active galactic nuclei, cosmic ray experiments, deep learning,
convolutional neural network, simulations}
\maketitle
\flushbottom

\section{Introduction}

Origin of the cosmic rays  of the highest energies ($E\gtrsim50$~EeV,
ultra-high-energy cosmic rays, \uhecrs) remains one of the biggest mysteries of the  high-energy astrophysics. Interaction with the photon backround limits the distance of CR
propagation to $\sim$100~Mpc at these energies.
It is natural to expect that CR sources trace to a certain degree the non-uniform
local matter distribution, and thus the distribution of \uhecr{} arrival
directions should be anisotropic.
An investigation of this anisotropy and its properties is one of the main avenues
in the \uhecr{} research as it could strongly constrain  the  characteristics of their  sources.

In this paper, we continue our studies of methods for probing a large-scale anisotropy arising
from a presence of a nearby source, which we began in~\cite{we-jcap2019,we-jcap2020}
(Papers~1 and~2 in what follows).
As a  representative of a broader class of models we used a particular model
for cosmic rays by Kachelrie\ss, Kalashev, Ostapchenko and Semikoz (KKOS
in what follows) \cite{Kachelriess:2017tvs}.  The model assumes that \uhecrs{} are accelerated by (a
subclass of) active galactic nuclei (AGN) with the energy spectra of
nuclei following a power-law with a rigidity-dependent cut-off after the
acceleration phase.  The model successfully reproduces the energy
spectrum of cosmic rays with energies beyond $10^{17}$~eV registered
with the Pierre Auger Observatory and the spectrum of high-energy
neutrinos registered by IceCube, as well as data on the depth of maximum
of air showers~\Xmax{} and RMS(\Xmax).  One of the consequences of the
model is the existence of a nearby (within $\sim20$~Mpc) AGN acting as a
source of UHECRs.  Presence of such an accelerator would inevitably lead
to deviations from isotropy at some level.   

In Paper~1, we investigated imprints that this deviation would leave on the  angular power spectrum (APS) of the UHECR flux depending on the contribution from the strongest individual source. We assumed that the source of \uhecrs{} is one of five close AGNs (Cen A, M82, M87, Fornax A, NGC 253), which are the most probable candidates in our scenario.
We demonstrated that an observation of
$\gtrsim200\text{--}300$ events with energies $\gtrsim57$~EeV would be necessary to reject the
hypothesis of isotropy  with a high level of statistical
significance if the fraction of events from any of these sources is
$\simeq$10--15\% of the total flux.

In our next paper, we decided to develop another approach to an analysis of
anisotropy because the APS calculation mostly destroys information about the 
characteristic shape and size of a region with events coming from a source, thus somehow decreasing  sensitivity to a sought signal. In Paper~2, we exploited full
information about arrival directions of \uhecrs{} by using methods of machine learning,
namely, a convolutional neural network (CNN) classifier. As arrival directions were  distributed across the celestial sphere, the CNNs were trained on a  HEALPix grid~\cite{healpix}.  This approach considerably increased sensitivity, as now the number of events needed for establishing a
large-scale anisotropy produced by \uhecrs{} arriving from a nearby
source was $\sim4$ times smaller than in the case of the traditional APS analysis.
For a fixed sample size, the CNN strongly decreased the fraction of CRs arriving
from the source necessary for a robust detection of an anisotropy.
It is important to mention that the KKOS model provides a heavy mass composition
of \uhecrs{}
at energies above 57~EeV thus resulting in much more fuzzy patterns of
arrival directions if compared with the case of a light (mostly proton and
helium) composition.
In this sense, our results were conservative since having more compact
patterns will allow obtaining less restrictive demands on the
minimal number of from-source events needed to reject the isotropy
hypothesis.

However, we only used information about arrival directions of UHECRs while there is
another observable, energy. Besides this, we did not take into account the intrinsic
uncertainties in the energy of primary cosmic rays arriving to Earth, which made
samples used for training and testing the CNNs more accurate than in any experiment.
Less energetic particles would be deflected by larger angles generally, populating
the periphery of the region filled with CRs arriving from a particular nearby source.
It is natural to expect that taking this uncertainty into account would diminish
our results making them less relevant for experimental data.
This motivated us to to make the next step in the development of a deep-learning-based
approach to anisotropy studies of \uhecrs{} by incorporating information
about energies and taking into account the above mentioned uncertainty.
The preliminary results of this study are presented in the letter.

\section{Method}
\label{sec:method}

Our analysis employs mock maps of \uhecr{} arrival directions that we generate
using the KKOS scenario as the starting point. We considered five nearby active galactic
nuclei Centaurus~A, M82,
NGC~253, M87 and Fornax~A as possible sources of \uhecrs. In our scenario, all of them share the same
injection spectrum and composition. However, these properties evolve during propagation of
nuclei from the source to the Milky Way, and we take this into account by using the TransportCR
code~\cite{transportcr}, considering only \uhecrs{} of the  energies above 56~EeV. This allowed
us to neglect deflections in extragalactic magnetic fields~\cite{Pshirkov:2015tua} and
to assume that \uhecrs{} arrive to the Milky Way within $\pm1^{\circ}$ from the source position.
Propagation from the Galactic boundary to the Earth was simulated using the CRPropa~3
code~\cite{CRPropa}.  The
calculations were performed on the HEALPix grid with $\Nside=512$. This allowed us to map
accurately arrival directions of \uhecrs{} of different rigidities at the Galactic boundary
to arrival directions observed at Earth using the method of backpropagation.

A mock map with $N$ events was generated as follows.
One takes the propagated spectrum at the Galactic boundary
calculated with TransportCR for a source located at a given distance from the Galaxy and samples it~$\Nsrc$ times, each time extracting some nuclei with an energy~$E$ and charge~$Z$.  An observed arrival direction
of a cosmic ray is found then for each $(E,Z)$ pair (or, equivalently, for each rigidity) using the mapping obtained with CRPropa by backpropagation.  The remaining $N-\Nsrc$ events are generated
following the isotropic distribution. The whole process is repeated many times in order to generate a sufficiently large number of maps for each source.

In Paper~1, we analysed these mock maps by calculating the angular power spectrum
of CR arrival directions 
following a method suggested by the IceCube and the Pierre Auger Observatory
collaborations~\cite{IceCube-aniso-2007,Auger-aniso-2017}. The APS was calculated
using maps of the relative intensity of the CR flux obtained from these mock maps.
After that, we calculated an estimator that quantified difference between our APS
and an APS of the isotropic flux.
It was assumed as the null
hypothesis~$H_0$ that arrival directions of a mixed sample of \uhecrs{}
obey an isotropic distribution.  We adopted the value of the error of
the second kind (the probability not to reject the null hypothesis when
it is false) $\beta=0.05$ and searched for a minimal fraction~$\eta$ of
from-source events in the total flux such that the error of the first
kind (the probability to reject the true null hypothesis)
$\alpha\lesssim0.01$. 


Eventually, the problem of source identification could be reformulated as as pattern
recognition task. Nowadays, tasks of this kind are often solved using Convolutional
Neural Networks (CNNs), a widely used subclass of feed-forward neural networks~\cite{CNN}.
We applied this approach for anisotropy analysis in our Paper~2.
CNNs use local feature maps at different 
scales to extract valuable information and perform a classification
task. In the simplest case, it answers a question whether a map belongs
to an ``isotropic'' or ``with source'' class.
CNNs are routinely implemented for many programming
languages and platforms, but mostly for flat images only.
In cosmic ray physics, one needs to study an ``image''
on a sphere, so these implementations could not be used straightforwardly.
Some implementations of spherical CNNs were proposed
recently~\cite{SphericalCNN,Perraudin:2018rbt,Krachmalnicoff:2019zjh}.
We employed the publicly available code developed by Krachmalnicoff and
Tomasi~\cite{Krachmalnicoff:2019zjh} with minor additions.  This code
implements the convolution and pooling (down-sampling) operations on
the HEALPix grid data with the help of Keras deep learning
library~\cite{keras}. The convolution operation on the HEALPix grid is
parameterized by 9 adjustable weights per feature map.
The CNN developed in Paper~2 takes 1 feature map in the HEALPix
grid with $\Nside=32$ as an input.\footnote{%
    $\Nside=32$ corresponds to  the sphere  divided into 12,288 cells with the angular size
    of $1.83^\circ$, which is of the order of the angular resolution of UHECR experiments.}


In the present work, we developed a natural extension of the above approach by increasing the number of feature maps analyzed by the convolutional neural network.
We split the energy range of interest to bins of size~$\Delta_b$ in $\lg E$ scale
and calculate expected event density maps for each energy bin before sampling individual events.
The from-source event density map calculation procedure is slightly different from the one used in Paper II. As before we take the propagated spectrum calculated with TransportCR for a source located at a given distance
from the Galaxy and sample it~$N_\mathrm{ini}=100,000$ times, each time extracting some
nuclei with an energy~$E$ and charge~$Z$. An observed arrival direction
of a cosmic ray is then found for each $(E,Z)$ pair using the previously calculated mapping
obtained with CRPropa code by backpropagation. Due to the uncertainty in energy determination,
each sampled $(E,Z)$ pair contributes to several neighbour energy bins with a weight given by
the normal distribution integral taken with the limits defined by the bin boundaries:
$$
w_i = \int_{\lg E_i}^{\lg E_i + \Delta_b} N(E, \Sigma_{\lg E})\,,
$$
where $\Sigma_{\lg E}$ is proportional to the relative energy determination uncertainty:
$$
\Sigma_{\lg E} \simeq \frac{1}{\ln 10} \frac{\Delta E}{E}.
$$
Maps generated this way are used then to sample~$\Nsrc$ from-source events,
characterized by HEALPix angular coordinates and energy bin. The remaining $N-\Nsrc$ events
are generated following the isotropic distribution assuming the same model energy spectrum,
with the sampling procedure accounting for the energy determination error in the same way.
The event sampling process is repeated multiple times as it was done before in order
to generate a sufficiently large number of maps for each source. Note that although
we impose the lower cut $E>56$~EeV on the observed event energies as
in Paper~2, this time we self-consistently account for the influence of the lower
energy part of the model on the analysis because of the non-vanishing energy
determination uncertainty. This is one of the major advantages of the new approach.
We assume the value of $20$\% for the energy determination uncertainty $\Delta E/E$,
which follows an estimate for the POEMMA mission~\cite{Olinto:2020oky}.

\begin{figure}[!t]
	\centering
	\figh{.5}{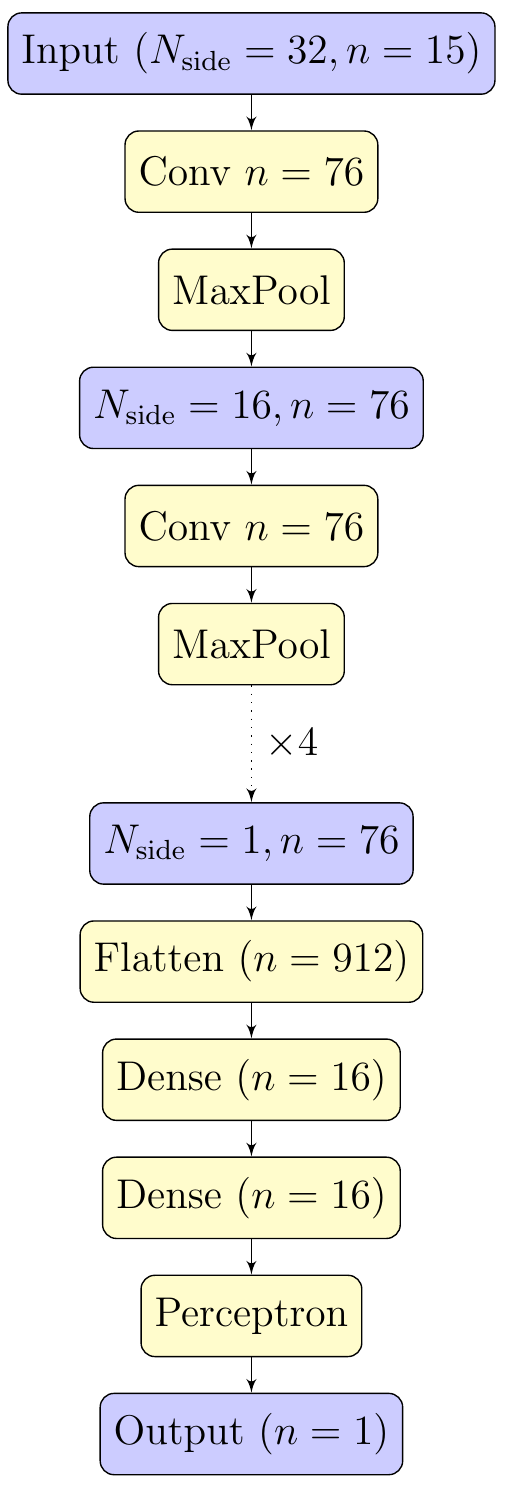}
	\caption{Architecture of the CNN developed in this work.
	Blue boxes are used to show feature vectors and maps. Yellow boxes
	show operations.}
	\label{fig:scheme}
\end{figure}

The classifier neural network architecture has been modified to take as an input
a collection of maps instead of single map. We also had to repeat the model
meta-parameter optimization procedure. As before, we tried to adjust the HEALPix
grid size along with number of filters and convolutions. Besides this, we found it
useful to add two extra dense layers after the sequence of convolutions.
The log energy bin size $\Delta_b$ was also one of the free parameters being optimized.
Among the three values probed, $1/5$, $1/10$ and $1/20$, the smallest one was found to
be optimal, which results in~15 energy bins in the range $56 < E/\text{EeV} < 315$~EeV.
The classifier neural network architecture which we found optimal is shown in figure~\ref{fig:scheme}.

The CNN developed in this work takes 15 feature maps in the HEALPix
grid with $\Nside=32$ as an input. 76 feature maps are built at the first step using the convolution operation with $76\times9\times15$ free parameters and max-pooling the image to $\Nside=16$.
The sequence of convolutions and max-pooling operations is repeated
until reaching $\Nside=1$ with the persistent number of feature maps,
which means that each intermediate convolution operation has
$76\times76\times9$ trainable weights.  The rectified linear activation
function (ReLU) is used for all intermediate layers.  Finally, 76
feature maps with $\Nside=1$ are flattened and processed by a multilayer-layer
perceptron with two inner layers of size 16 followed by single neuron output with sigmoid activation.

To avoid overfitting, we used an early-stop
technique.  Namely, we trained our model for at most 100 epochs with 10,000 map samples per epoch and interrupted training in case accuracy on validation data was not improving for 10 epochs. Each batch of training set was generated randomly which helped to prevent ovefitting. For this reason we did not use any other regularization techniques, such as dropout and the L2 regularization. The validation set consisted of 10,000 samples generated using different random seed. 
The weights were optimized using the Adadelta adaptive learning rate
method~\cite{adadelta}. Finally we evaluated our model on the test set consisting of 50,000 map samples. The output of the classifier, a number between~0 and~1 was used as the test statistic.

The minimal fractions~$\eta$ of from-source \uhecrs{} needed
to reject the null hypothesis of an isotropic flux with the same demands
on~$\alpha$ and~$\beta$ as above, are presented in
table~\ref{table:percentNN}.
Results obtained in Paper~2 are included for comparison.
It is clearly seen that opposite to the expectations, the results
obtained for the binned classifier are essentially the same as for
the CNN developed earlier despite of the more fuzzy patterns of
arrival directions of \uhecrs{} coming from a particular source.

\begin{table}[!t]

	\caption{Percentage of UHECRs arriving from the candidate sources
		in samples of sizes $N=50, 100,\dots,500$ such that the
		error of the first kind
		$\alpha\lesssim0.01$ for the null hypothesis of isotropy~$H_0$
		providing the second kind error $\beta=0.05$, obtained
		with the new binned convolutional neural network classifier (BCNN) shown
		in figure~\ref{fig:scheme} and with the previous convolutional neural
		network (CNN) model. One source at a time was considered.}

	\label{table:percentNN}
	\medskip
	\centering
	\begin{tabular}{|l|c|c|c|c|c|c|c|}
		\hline
		Source    & Method &\,\,50\,\,&100&200&300&400&500\\
		\hline
		NGC 253   &BCNN& 12 & 8 & 5 & 4 & 3.25  & 3.2 \\
					 &CNN& 12 & 7 & 4.5 & 3.67 & 3 & 2.6 \\ 
		\hline
		Cen A     &BCNN& 16 & 10 &  7 &  5.67  & 4.75 &   4.2 \\
					 &CNN& 16 & 11 & 7  & 5.67&  5 & 4.4 \\
		\hline
		M 82      &BCNN& 20 & 14 &  8.5 &  6.67  & 5 &   4.4 \\
					 &CNN& 20 & 12 &  7 & 6  & 4.75 & 4.2 \\
		\hline
		M 87      &BCNN& 22 & 15 &  11.5 &  7  & 6.25 &   5.4 \\
					 &CNN& 22 & 14 & 9  &  8 & 6.25& 5.2 \\  
		\hline
		Fornax A  &BCNN& 16 & 10 &  6 &  5  & 4.5 &   3.8 \\
					 &CNN& 16 &  9 &  6 &  5 & 4.5 & 3.8 \\
    	\hline
	\end{tabular}
\end{table}

\section{Conclusions}

We demonstrated in Paper~2 how one can strongly improve the efficiency of an
analysis of arrival directions of \uhecrs{} by introducing a deep
convolutional neural network trained on a HEALPix grid.
The basic idea was to train a classifier which discriminates
samples generated assuming null (isotropy) and alternative (anisotropy)
hypotheses and to use the classifier output as a test statistic.
An application of models that involve pattern recognition such as the
suggested CNN gave a
qualitative enhancement in terms of sensitivity to deviations
from an isotropic distribution of arrival directions. It was shown in
particular that the method allows decreasing the minimal number of
events necessary to reject the null hypothesis by $\sim4$ times.
This reduces technical demands and the required total exposure
of an UHECR experiment drastically.

However, the model considered was a simplified one since it
did not take into account uncertainties in the energy
of detected \uhecrs, which influence the spectrum and thus the shape
of patterns formed by nuclei arriving from a source.
As a result, our estimates could be too optimistic.
We addressed the problem in the present work by introducing a binned
classifier by modifying the neural network developed in Paper~2.
It was demonstrated that even under conditions of a 20\% uncertainty
in the energy resolution (which is higher than in contemporary
ground experiments), the quality of the new classifier did not decrease.
More than this, an introduction of energy as a new observable opens a
way for an improvement of the efficacy of discriminating samples
generated assuming isotropy and anisotropy hypotheses.
This will be studied in more details elsewhere.

The source code and supplemental materials for this work,
including trained classifier models, can be downloaded from the project
web page~\cite{gitrepo}.

\acknowledgments

The research has made use of the NASA/IPAC Extragalactic Database (NED),
which is operated by the Jet Propulsion Laboratory, California Institute
of Technology, under contract with the National Aeronautics and Space
Administration, and of the SIMBAD database, operated at CDS, Strasbourg,
France~\cite{simbad}.
Some of the results in this paper were obtained using the
HEALPix package~\cite{healpix}. 
The development of the classification method and
the architecture of the corresponding deep convolutional neural network
is supported by the Russian Science Foundation grant 17-72-20291.
The authors acknowledge support of the Interdisciplinary Scientific and
Educational School of Lomonosov Moscow State University
``Fundamental and Applied Space Research.''

\bibliographystyle{JHEP}
\bibliography{nnaniso}

\end{document}